\documentclass[debug]{rmaa}

\title{Metallicity effects on the modified wind momentum of CSPN} 

\author{
  W. J. Maciel,\altaffilmark{1} 
  G. R. Keller,\altaffilmark{1}
  and R. D. D. Costa\altaffilmark{1}}

\altaffiltext{1}{Instituto de Astronomia, Geof\'\i sica e Ci\^encias
                 Atmosf\'ericas, Universidade de S\~ao Paulo, Brazil.}

\shortauthor{Maciel, Keller, \& Costa}
\shorttitle{Metallicity effects in CSPN winds}

\fulladdresses{
\item W. J. Maciel, G. R. Keller and R. D. D. Costa: 
    Instituto de Astronomia, Geof\'\i sica e Ci\^encias
    Atmosf\'ericas, Universidade de S\~ao Paulo - 
    Rua do Mat\~ao 1226, CEP 05508-900, S\~ao Paulo SP, Brazil
    (maciel@astro.iag.usp.br, graziela@astro.iag.usp.br,
    roberto@astro.iag.usp.br).}

\listofauthors{W. J. Maciel, G. R. Keller, \& R. D. D. Costa}
\indexauthor{Maciel, W. J.}
\indexauthor{Keller, G. R.}
\indexauthor{Costa, R. D. D.}

\abstract{Recent investigations on the central stars of planetary nebulae
(CSPN) indicate that the masses based on model atmospheres can be much 
larger than the masses derived from theoretical mass-luminosity relations.
Also, the dispersion in the relation between the modified wind momentum 
and the luminosity depends on the mass spread of the CSPN, and is larger 
than observed in massive hot stars. Since the wind characteristics probably 
depend on the metallicity, we analyze the effects on the modified wind 
momentum by considering the dispersion in this quantity caused 
by the stellar metallicity. Our CSPN masses are based on a relation 
between the core mass and the nebular abundances. We conclude that 
these masses agree with the known mass distribution both for CSPN and white 
dwarfs, and that the spread in the modified wind momentum can be explained 
by the observed metallicity variations.}

\resumen{ }

\addkeyword{Planetary nebulae}
\addkeyword{Stars: Planetary nebula central stars}
\addkeyword{Stars: Mass loss}

\begin{document}
\maketitle

\section{Introduction}
\label{section1}

Stellar winds are observed in practically all the HR diagram, from dwarf, 
solar-like stars to the hot supergiants and massive stars in the main sequence. 
In particular, AGB stars have slow, massive winds leading to the ejection of 
the stellar outer layers as a planetary nebula (PN). At this stage, the stellar 
remnant is extremely hot, originating a fast, massive wind which is similar to 
the radiative winds observed in hot, massive young stars. These winds are observed 
in central stars of planetary nebulae (CSPN), both for H-rich stars and for stars 
with Wolf-Rayet characteristics. The origin of the winds of these stars is 
attributed to the stellar radiation pressure on metallic lines, a similar 
mechanism known to be active in hot, massive galactic stars.

In the last few years, some work has been done in order to use the available 
information on the stellar winds of CSPN to determine physical properties of 
the stars, such as their radii, luminosities and masses (see for example Kudritzki
\& Puls \citeyear{kp2000}, Pauldrach et al. \citeyear{pauldrach03}). In particular, 
observed or inferred properties such as the mass loss rate and the wind terminal 
velocity can be used in order to constrain some physical properties of the stars, 
such as their luminosities and masses. As a result, there seems to exist a discrepancy 
between recent mass determinations for CSPN based on model atmospheres affected by 
winds and the corresponding determinations from mass-luminosity 
relations derived from evolutionary tracks of post-AGB stars. There is a much 
larger mass spread obtained from model atmospheres as compared with the predictions 
of standard mass-luminosity relations and with the known mass distribution of 
CSPN and white dwarfs. 

For a stellar wind with mass loss rate $\dot M = dM/dt$ and terminal velocity 
$v_\infty$ in a star with radius $R_\ast$, a correlation can be obtained between 
the modified wind momentum, defined as $p_w = \dot M\, v_\infty \, \sqrt{R_\ast}$ 
and the stellar luminosity $L/L_\odot$. This correlation is well defined for 
massive stars, and is also  approximately valid for CSPN. However, use of a 
standard mass-luminosity relation implies a relatively large spread in the 
modified wind momentum at a given luminosity for CSPN, in contrast with the 
results from model atmospheres, which show a better agreement with the relation 
derived from massive stars (Kudritzki et al. \citeyear{kudritzki97}; Pauldrach
et al. \citeyear{pauldrach04}).

In the present work, we analyze the available information on the wind momentum in 
a sample of well-studied CSPN in the Galaxy, and use this information in order to 
constrain basic stellar properties. In particular, we look for correlations involving 
the observed wind and nebular properties, especially the chemical abundances. In fact, 
the stellar metallicity (and as a consequence the nebular chemical composition) has 
some effect on the wind properties, which can be seen for example by comparing galactic 
stellar winds with the corresponding quantities in stars of the Magellanic Clouds, 
which are more metal-poor than the Milky Way. This can be explained by the fact that 
the radiative mechanism operates essentially on the metal absorption lines, so that 
the stellar metallicity plays a role in the mass loss  process. On the other hand, 
there is a clear correlation between the nebular chemical composition and the stellar 
mass, especially regarding those element ratios that are affected by the evolution of 
the PN progenitor stars, such as the N/O and He/H ratios (cf. Maciel \citeyear{maciel00}). 
As a consequence, the observed wind properties, such as the wind momentum or some related 
quantity should in principle be related to stellar mass and metallicity. The IAG/USP group 
has a considerable experience in the determination and analysis of the chemical composition 
of PN in the Galaxy and the Magellanic Clouds (see for example Costa et al. \citeyear{costa2004})
and references therein. In this work we use this database and examine the effects
of the nebular metallicity in the observed properties of the corresponding stellar winds
in a sample of well studied galactic CSPN.

\section{Masses of CSPN}
\label{section2}

\subsection{Previous mass determinations}

As recently discussed by Pauldrach et al. (\citeyear{pauldrach03}, \citeyear{pauldrach04}), 
there is a clear discrepancy between the mass-luminosity relation for CSPN as derived from 
their improved model atmospheres and the relation previously obtained by Kudritzki et al. 
(\citeyear{kudritzki97}), which was obtained from the application of post-AGB 
evolutionary tracks. The new results produce a larger spread in the stellar masses, 
$0.4 < M/M_\odot < 1.4$, as compared with the range $0.6 < M/M_\odot < 1$ from Kudritzki
et al. (\citeyear{kudritzki97}). Moreover, the predicted luminosities are lower
for a given mass in the new models, and the overall correlation is less well defined,
as can be seen from Fig.~1. In this figure, empty circles are the results by
Kudritzki et al. (\citeyear{kudritzki97}) and the filled circles refer to the
data by Pauldrach et al. (\citeyear{pauldrach04}). The objects included in the figure
are NGC 2392, NGC 3242, IC 4637, IC 4593, He2-108, IC 418, Tc 1, He2-131, and NGC 6826.
The values of the corresponding quantities (stellar radius, mass, luminosity,
terminal velocity, mass loss rate, etc.) are given in the original papers by
Kudritzki et al. (\citeyear{kudritzki97}) and Pauldrach et al. (\citeyear{pauldrach03};
\citeyear{pauldrach04}).


   \begin{figure*}
   \centering
   \includegraphics[width=12.0cm]{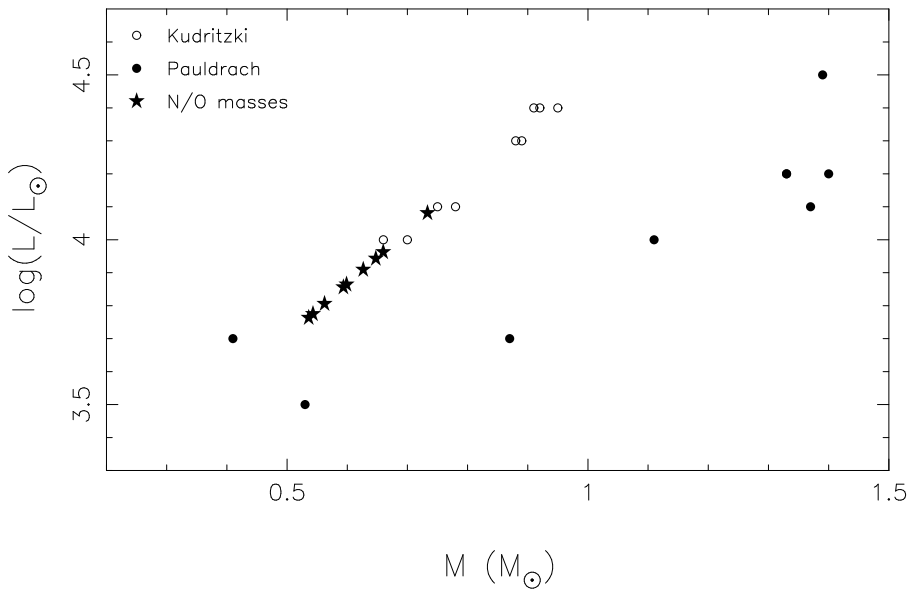}
      \caption{Luminosities and masses of CSPN. Empty circles: data by
      Kudritzki et al. (\citeyear{kudritzki97}), filled circles: data
      by Pauldrach et al. (\citeyear{pauldrach04}); stars: N/O masses
      (see text).}
   \label{fig1}
   \end{figure*}

The work of Pauldrach et al. (\citeyear{pauldrach03}; \citeyear{pauldrach04})
is based on hydrodynamically consistent, spherically symmetric model atmospheres
which are able to reproduce the observed ultraviolet spectra of hot, massive
stars. On the basis of the observed wind characteristics, namely the terminal
velocity and the mass loss rate, the main stellar parameters can be derived.
In this case, the obtained stellar mass is independent of any adopted mass-luminosity
relation, which in principle may be an advantage. On the other hand, the results
by Kudritzki et al. (\citeyear{kudritzki97}) are based on a core mass-luminosity
relationship for post-AGB stars, in the framework of a modelling of stellar H
and He line profiles. Although this method may be affected by complications such
as a contamination by nebular lines, it generally provides accurate determinations
of the wind properties as well as of the stellar gravity and mass. Therefore,
we need further information in order to clarify the discrepancy in the stellar
masses observed in Fig.~1.

\subsection{The N/O masses}

This problem can be investigated taking into account the CSPN masses derived from 
observed nebular abundances of N/O, as discussed in Cazetta \& Maciel (\citeyear{cazetta})
and Maciel (\citeyear{maciel00}). From AGB evolution theory, there is a general 
correlation between the N/O abundance ratio and the core mass,  that is, the CSPN mass 
(see for example Marigo \citeyear{marigo}). Theoretical models for AGB stars predict 
that higher stellar masses are associated with larger N/O abundances, which 
are enhanced by the dredge-up episodes that occur in the CSPN, especially  the second one. 
Cazetta \& Maciel (\citeyear{cazetta}) and Maciel (\citeyear{maciel00}) presented a 
detailed discussion of the core mass-N/O abundance relation based both on
theoretical models of AGB stars and on empirical mass determinations. This relation
takes into account several initial mass-final mass relations available in the literature, 
as well as different individual mass determinations, so that it can be considered
as independent of a single  mass-luminosity relation, such as the one adopted by
Kudritzki et al. (\citeyear{kudritzki97}). As a result, two calibrations were 
determined, which were referred to as the high-mass and the low-mass calibration
(cf. Maciel \citeyear{maciel00} for details). The latter is considered as more accurate,
as it produces core masses $M_\ast \geq 0.55\,M_\odot$ and main sequence masses
$M_{MS} \geq 1\,M_\odot$, in agreement with  the detailed CSPN masses of 
Stasi\'nska et al. (\citeyear{stasinska}) as well as with the masses of Type~II PN
proposed by Peimbert (\citeyear{peimbert}). The adopted initial mass-final mass
relation was based on the gravity distance work of Maciel \& Cazetta (\citeyear{mc97}).

It is then interesting to investigate whether the masses obtained from the 
analysis of nebular abundances can be used to distinguish between the masses derived 
from model atmospheres and stellar wind analysis. 

Following Cazetta \& Maciel (\citeyear{cazetta}) and Maciel (\citeyear{maciel00}), 
the core mass (in solar masses) can be written as

   \begin{equation}
      M_\ast = 0.7242 + 0.1742 \ \log{\rm (N/O)}  \ ,
      \label{mno1}
   \end{equation}

\noindent
for $-1.2 \leq  \log{\rm (N/O)} < -0.26$ and

   \begin{equation}
      M_\ast = 0.825 + 0.936 \ \log{\rm (N/O)} + 1.439 \ [\log{\rm (N/O)}]^2 \ ,
      \label{mno2}
   \end{equation}
 
\noindent
for $-0.26 \leq  \log{\rm (N/O)} < 0.20$, where (N/O) is the nitrogen over 
oxygen abundance ratio by number of atoms. Columns 3 to 6 of Table~1 show the 
abundances of oxygen, nitrogen, and N/O adopted from our IAG/USP database, and the 
calculated masses for the sample of CSPN shown in Fig.~1. The object IC~4637 is not 
in our database, so that we have adopted an average abundance $\log({\rm N/O}) = 
-0.56$, which is typical for elliptical planetary nebulae, according to the discussion 
of G\'orny et al. (\citeyear{gorny}). 
 
\begin{table}
\caption[]{Data for the central stars.}
\label{table1}
\begin{flushleft}
\begin{tabular}{llcccccc}
\noalign{\smallskip}
\hline\noalign{\smallskip}
 & Name& $\epsilon$(O) & $\epsilon$(N)& log(N/O) & $M_\ast (M_\odot)$ 
& $\log(L_\ast/L_\odot)$ & $\log p_w$ \\
\noalign{\smallskip}
\hline\noalign{\smallskip}
1 &  NGC 2392 &   8.50 & 8.38 & -0.12 & 0.73 & 4.08 & 26.63 \\
2 &  NGC 3242 &   8.66 & 7.91 & -0.75 & 0.59 & 3.85 & 26.44 \\
3 &  IC 4637  &        &      & -0.56 & 0.63 & 3.91 &       \\
4 &  IC 4593  &   8.32 & 7.24 & -1.08 & 0.54 & 3.77 & 25.98 \\
5 &  He2-108  &   8.22 & 7.85 & -0.37 & 0.66 & 3.96 & 26.17 \\
6 &  IC 418   &   8.54 & 7.82 & -0.72 & 0.60 & 3.87 & 26.35 \\
7 &  Tc 1     &   8.71 & 7.67 & -1.04 & 0.54 & 3.77 & 26.37 \\
8 &  He2-131  &   8.67 & 8.23 & -0.44 & 0.65 & 3.95 & 26.60 \\
9 &  NGC 6826 &   8.43 & 7.50 & -0.93 & 0.56 & 3.80 & 26.14 \\
\noalign{\smallskip}
\hline
\end{tabular}
\end{flushleft}
\end{table}

Fig.~2 shows the masses as given by Pauldrach et al. (\citeyear{pauldrach04}, solid dots) 
and Kudritzki et al. (\citeyear{kudritzki97}, empty circles) as a function of the masses 
given in Table~1, which we may call N/O masses. It can be seen that the N/O masses are 
generally closer to the masses derived by Kudritzki et al. (\citeyear{kudritzki97}) 
and Kudritzki et al.  (\citeyear{kudritzki06}) than to the values by Pauldrach et al. 
(\citeyear{pauldrach04}), which show a much higher mass dispersion. In view of the
similarity between our N/O masses and the most of the masses obtained by Kudritzki et al. 
(\citeyear{kudritzki97}, it is tempting to calibrate the luminosities using the 
mass-luminosity relation used in that work, which can be approximately written as

   \begin{equation}
      \log (L_\ast/L_\odot) \simeq 2.90 + 1.61 \, (M_\ast/M_\odot) \ .
      \label{mlr}
   \end{equation}

   \begin{figure*}
   \centering
   \includegraphics[width=12.0cm]{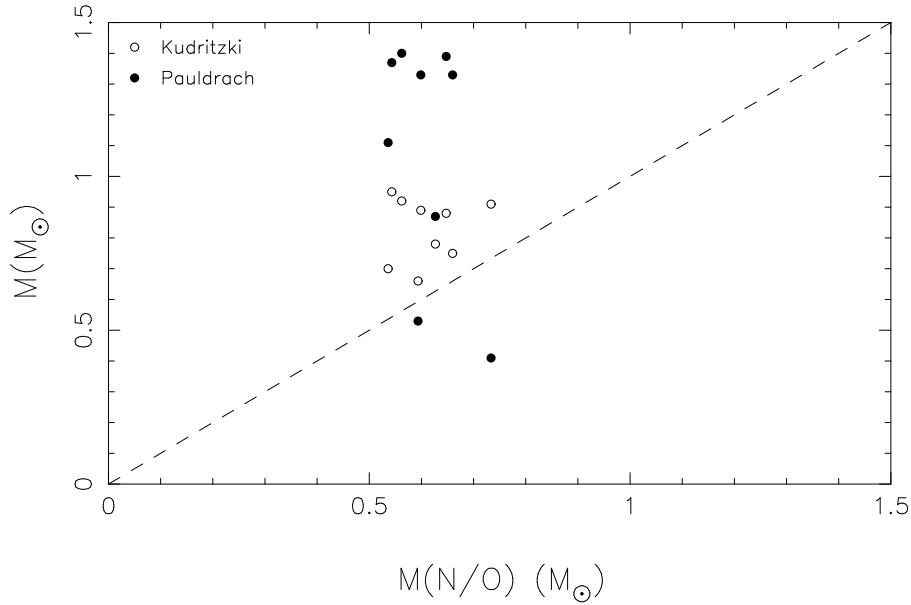}
      \caption{Masses of CSPN as a function of the N/O masses given
      in Table~1. Empty circles: Kudritzki et al. (\citeyear{kudritzki97}), 
       filled circles: Pauldrach et al. (\citeyear{pauldrach04}).}
   \label{fig2}
   \end{figure*}

\noindent
The derived luminosities are given in column~7 of Table~1 and the N/O masses
and corresponding luminosities are also included in Fig.~1 as black stars. It can be
seen that these masses have a much more limited range, roughly $0.5 < M (M_\odot) < 0.8$, 
than obtained by Pauldrach et al. (\citeyear{pauldrach04}), and are also somewhat more
restricted than the values by Kudritzki et al. (\citeyear{kudritzki97}). 

\subsection{Discussion}

A comparison of the CSPN masses discussed in the previous subsections can be made
both on the basis of the known mass distributions of these objects and their white
dwarf descendants as well as of the rather limited individual mass determinations
for the central stars.

A set of central star masses was obtained by G\'orny et al. (\citeyear{gorny}),
based on measured nebular expansion velocities and central star parameters,
within the framework of a simple evolutionary model for planetary nebulae.
A comparison of our derived masses given in Table~1 with their results shows 
a good agreement, as all objects have a mass difference under 0.10\,$M_\odot$,
and five of the stars have even smaller differences, under 0.03\,$M_\odot$. 
The only exception is NGC 2392, for which our derived mass is 0.12\,$M_\odot$ 
larger than the mass given by  them, a difference of about 17\%. However, this 
object has a very high nitrogen (and He, see Section~3) abundance, which is 
consistent with a larger mass than derived by G\'orny et al. (\citeyear{gorny}).

Recently, Zijlstra et al. (\citeyear{zijlstra}) presented the results of a 
25-year monitoring program of radio observations of the planetary nebula
NGC 7027, possibly the best studied object of its kind. From the evolution
of the radio flux during this period, an improved expansion distance was
obtained. Taking into account theoretical tracks for post-AGB stars, a
distance-independent mass of $0.655\pm 0.01\,M_\odot$ was derived. This
result agrees very well with our N/O mass for this object, which is 
$0.66\,M_\odot$, adopting abundances from our database (see for example
Maciel \& Quireza \citeyear{mq99}, Maciel \& Chiappini \citeyear{mc94}).
Using alternative models, Zijlstra et al. (\citeyear{zijlstra}) find
a slightly larger mass, still in good agreement with our result.

Also, in a recent work Traulsen et al. (\citeyear{traulsen}) analyzed high
resolution ultraviolet spectra of a few hydrogen-rich CSPN obtained with HST and 
FUSE, and were able to locate these objects in a $\log g \times \log T_{eff}$
plane using theoretical evolutionary tracks. For the three objects in our
abundance dataset (NGC  4361, NGC 6853, and NGC 7293, see also Perinotto et al. 
\citeyear{perinotto04} and Henry et al. \citeyear{henry04}), the central star masses
obtained by Traulsen et al. (\citeyear{traulsen}) cluster around $0.6\,M_\odot$,  
within about 15\% of the N/O masses derived for these objects.

Stasi\'nska et al. (\citeyear{stasinska}) applied the method described by
G\'orny et al. (\citeyear{gorny}) and derived a mass distribution of CSPN
with a very restricted range, in which more than 80\% of the objects have 
masses between $0.55$ and $0.65\,M_\odot$. This distribution peaks around
$0.60\,M_\odot$ with some dispersion that depends on the adopted nebular mass,
and falls steeply towards large masses. Therefore, these results fit well the
lower masses found by the N/O method, so that the large masses found in a few
cases by Kudritzki et al. (\citeyear{kudritzki97}) and especially by
Pauldrach et al. (\citeyear{pauldrach04}) seem rather unlikely. A similar
conclusion has been recently reached by Napiwotzki (\citeyear{napiwotzki}),
who argued that the high CSPN masses close to the Chandrasekhar limit derived
by Pauldrach et al. (\citeyear{pauldrach04}) are physically implausible on the
basis of kinematical parameters extracted from galactic orbits and average
abundances for the different PN types.

Recent work on the much better known mass distribution of white dwarfs fully
support this conclusion, as can be seen from the white dwarf mass distribution of 
Madej et al. (\citeyear{madej}), especially regarding the largest mass observed in 
the sample, which in our case is under $0.8\,M_\odot$. According to the investigation 
by Madej et al. (\citeyear{madej}), which was based on a large sample of about 1200 
white dwarfs  from the Sloan Digital Sky Survey, their mass distribution peaks around 
$0.56\,M_\odot$, with very few objects having masses larger than about $0.8\,M_\odot$, 
in excellent agreement with our results. According to this work, less than 5\%
of the central stars are expected to have masses larger than $0.8\,M_\odot$,
a much lower fraction than observed either in the samples by  Kudritzki et al. 
(\citeyear{kudritzki97}) and Pauldrach et al. (\citeyear{pauldrach04}).

The mass distributions of CSPN and white dwarfs have also been recently considered
by Gesicki \& Zijlstra (\citeyear{gesicki}), based on a dynamical method which allows
mass determinations within $0.2\,M_\odot$. It results that both the CSPN and white
dwarf distributions peak around $0.6\,M_\odot$ as in Madej et al. (\citeyear{madej})
and Stasi\'nska et al. (\citeyear{stasinska}), although the white dwarf distribution
shows a broader mass range. The CSPN distribution has essentially no objects
with masses higher than $0.7\,M_\odot$ in their sample, while for the DA white
dwarfs -- presumably the progeny of H-rich CSPN -- a small fraction of objects
have masses larger than $0.8\,M_\odot$, similar to the results of Madej et al. 
(\citeyear{madej}). Although there may be some differences between the recently
obtained white dwarf mass distributions, as discussed by Gesicki \& Zijlstra
(\citeyear{gesicki}), they all agree in the sense that any sizable sample of 
CSPN is expected to have masses close to $0.60\,M_\odot$, a result that is 
reproduced by our N/O masses.

A further support to the N/O masses comes from the revision by Tinkler \& Lamers
(\citeyear{tinkler}) of the Kudritzki et al. (\citeyear{kudritzki97}) masses based
on a homogeneous set of parameters derived from Zanstra temperatures, dynamical
ages of the nebulae and evolutionary tracks. According to these results, an 
average mass of $0.60\,M_\odot$ is obtained, in excellent agreement with the 
results of Table~1, and showing no objects with masses larger than $0.8\,M_\odot$.
 
Finally, it could be mentioned that 5 central stars of PN in the galactic bulge
have been recently studied by Hultzsch et al. (\citeyear{hultzsch}) based on
high-resolution Keck spectra and NLTE modelling. In view of the fact that the 
distance to the galactic bulge is well known, the stellar parameters can be 
determined with better accuracy than in the case of field PN, although there
is some uncertainty due to the adopted extinction law. The masses derived
from their \lq\lq Method 1\rq\rq, which seems to be more reliable, come from
evolutionary tracks on the $\log T_{eff} \times \log g$ diagram, and are lower
than about $0.80\,M_\odot$ in all cases. Also, the estimated luminosities are close 
to the values obtained by Eq.~(3), supporting our results given in Table~1.

\section{The modified wind momentum of CSPN}
\label{section3}

For a star with radius $R_\ast$ having a wind with terminal velocity $v_\infty$
and a mass loss rate $dM/dt$, the modified wind momentum is defined as

   \begin{equation}
      p_w = {dM \over dt} \ v_\infty \ \sqrt{R_\ast} \ .
      \label{mwm}
   \end{equation}

It is well known from radiation driven wind theory that the mass loss rate is a function 
of the stellar luminosity $L_\ast$ and the effective mass $M_{eff}$, which can be written 
as

   \begin{equation}
    {dM \over dt} \propto L_\ast^{1/\alpha} \ M_{eff}^{{\alpha - 1 \over \alpha}} \ ,
      \label{dmdt}
   \end{equation}

\noindent
where $\alpha$ is the so-called force multiplier parameter, which is about $\alpha \simeq 2/3$ 
for hot stars (see for example Lamers \& Cassinelli \citeyear{lc99}; Pauldrach et al. 
\citeyear{pauldrach03}). The exponent $\alpha$ is the power law exponent of the line strength 
distribution function. The effective mass $M_{eff}$ is simply the stellar mass $M_\ast$ 
corrected for the electron scattering radiative force, $M_{eff} = M_\ast \ (1 - \Gamma_e)$,
where $\Gamma_e$ is the electron scattering efficiency. 

As a first approximation, $v_\infty$ can be assumed to be proportional to the stellar escape 
velocity $v_{esc} = (2 G M_{eff} / R_\ast)^{1/2}$, so that $v_\infty \propto 
\sqrt{M_{eff}/R_\ast}$. For example, for a large sample of galactic CSPN from a variety
of sources (Perinotto \citeyear{perinotto}; Kudritzki et al. \citeyear{kudritzki97};
Malkov \citeyear{malkov1}; Pauldrach et al. \citeyear{pauldrach04}), we obtain
a scaling factor between the terminal and escape velocity of about 2 to 4. Combining this 
equation with eqs.~(4) and (5), it is easy to see that the modified wind momentum depends 
weakly on the stellar mass, so that a relation of the form $p_w \propto L_\ast^{1/\alpha}$ 
is obtained. For instance, Pauldrach et al. (\citeyear{pauldrach03}, \citeyear{pauldrach04}) 
obtained a relation that produces a very good fit to the properties of hot massive galactic
stars given by

   \begin{equation}
     \log\, p_w = 20.479 + 1.507 \ \log(L_\ast/L_\odot) \ ,
     \label{pwl}
   \end{equation}

\noindent 
where $p_w$ is in g cm s$^{-2}$.  Fig.~3 shows the modified wind momentum with CSPN data by 
Kudritzki et al. (\citeyear{kudritzki97}) (empty circles) and Pauldrach et al. 
(\citeyear{pauldrach04}) (filled circles), using the same symbols as before, as a function 
of the stellar luminosity. The solid line represents the fit to hot, massive stars by 
Pauldrach et al. (\citeyear{pauldrach04}) given by Eq.~6. These stars are not shown in Fig.~3,
and occupy the upper right corner of the figure. Note that for NGC 2392, NGC 3242, 
IC 4637 and Tc 1 the mass loss rates from Kudritzki et al. (\citeyear{kudritzki97}) are 
upper limits.

Adopting the new N/O stellar masses, as shown in the previous section, and the luminosities 
from the mass-luminosity calibration by Kudritzki et al. (\citeyear{kudritzki97}), the 
stellar radii can be obtained, keeping the effective temperatures originally determined. 
Therefore, the modified wind momentum $p_w$ can be determined  using the originally derived 
$\dot M$ and $v_\infty$ and plotted as a function of the luminosity. This is also shown in 
Fig.~3, where the the open triangles correspond to our correction of the results by 
Kudritzki et al. (\citeyear{kudritzki97}) and filled triangles refer to $\dot M$ and 
$v_\infty$ data by Pauldrach et al. (\citeyear{pauldrach04}). It is clear that this is not 
a strictly consistent procedure, as the main stellar and wind parameters (effective temperature, 
gravity, abundances, radius, terminal velocity and mass loss rate) should in principle be 
derived in a consistent way. However, most of these parameters are essentially unchanged in 
both analyses. In fact, a direct comparison of the Kudritzki et al. and Pauldrach et al. 
results shows that there is a general agreement for $\log g$, $v_\infty$, $dM/dt$, $R_\ast$, 
and $T_{eff}$. Only the stellar mass and luminosity show larger discrepancies: the masses by 
Pauldrach et al. are generally larger than those by Kudritzki et al., while the opposite is 
true for the luminosities.

   \begin{figure*}
   \centering
   \includegraphics[width=12.0cm]{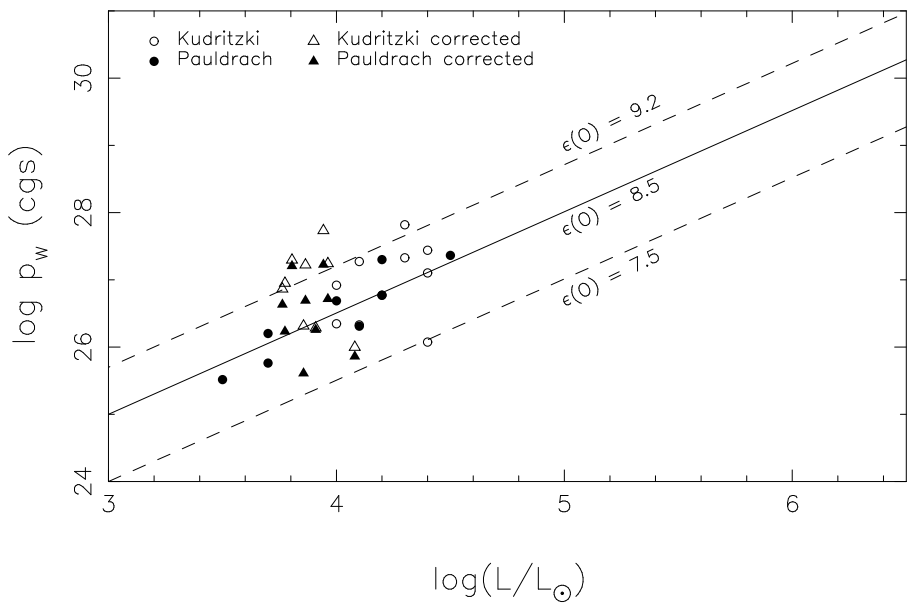}
      \caption{The modified wind momentum as a function of the stellar
      luminosity. Symbols are as in the previous figures: empty 
       circles: data by Kudritzki et al. (\citeyear{kudritzki97}); 
       filled circles: data by Pauldrach et al. (\citeyear{pauldrach04});
       emtpy and filled triangles are the corresponding corrected values 
       adopting the N/O masses. The metallicity dependence of the modified 
       wind momentum is illustrated by the dashed lines, corresponding
       to $\epsilon({\rm O})$ = 7.5 and 9.2 (see text).}
   \label{fig3}
   \end{figure*}

From Fig.~1 we can see that the new N/O masses of CSPN have a more restricted range, so that 
the same can be expected for the luminosities, as confirmed by Fig.~3. From Fig.~3 we can 
see also that the modified wind momentum has now a less tight correlation with the luminosity, 
similar to the results by Kudritzki et al. (\citeyear{kudritzki97}) and in disagreement with 
the results by Pauldrach et al. (\citeyear{pauldrach03}; \citeyear{pauldrach04}). It can also 
be seen that the same results are obtained for both sources of data, that is, the empty and 
filled triangles in Fig.~3 show essentially the same dispersion. As mentioned in Sect.~2,
these results are in agreement with the results by Napiwotzki (\citeyear{napiwotzki}), who 
has argued against the large masses for CSPN found by Pauldrach and co-workers, on the basis 
of the use of kinematic parameters obtained from galactic orbits and average abundances of 
the different PN types.

It should be mentioned that the procedure adopted here does not significantly affect the
stellar gravity $\log g$, which is a parameter generally well determined from the
observations. In fact, a detailed comparison of the resulting $\log g$ values with those
originally derived either by Kudritzki et al. (\citeyear{kudritzki97}) or Pauldrach et al. 
(\citeyear{pauldrach03}; \citeyear{pauldrach04}) shows an excellent agreement, in the
sense that (i) our values lie between the original values determined by these sources,
and (ii) the average deviations in $\log g$ amount to 0.2 and 0.1 dex for the 
Kudritzki et al. (\citeyear{kudritzki97}) and Pauldrach et al.  (\citeyear{pauldrach03}; 
\citeyear{pauldrach04}) data, respectively. An alternative approach would be to derive
the masses from the N/O abundances and the stellar radius from the original $\log g$
values, so that the luminosity would follow from the observed effective temperature.
However, we preferred  the procedure described earlier, since the mass-luminosity
relation is well founded theoretically, and any modifications in the adopted calibration
could be easily introduced in this procedure.

The N/O masses depend on the measured relative abundances of nitrogen and oxygen,
so that uncertainties in these measurements would affect the derived masses and
consequently, the luminosities. However, N and O abundances relative to hydrogen
are among the best determined abundances in photoionized nebulae, with uncertainties
generally lower than 0.2 dex. Abundances relative to oxygen, such as the N/O ratio
needed by our procedure, are generally better determined, usually within 0.1 dex.
Adopting this as an average, Eqs. (1) and (2) would lead to a maximum uncertainty
of about 10\% for central star masses in the range $0.6 < M/M_\odot < 0.8$. Clearly,
such an uncertainty would have a minor influence in the derived luminosities as 
plotted in Fig.~3.

\section{Metallicity dependence of the modified wind momentum of CSPN}
\label{section4}

\subsection{The modified wind momentum}

A key point in the analysis of the modified wind momentum is the expected dispersion in the 
plot of $\log p_w \times \log L/L_\odot$ for a given luminosity. The results by Kudritzki et al.  
(\citeyear{kudritzki97}) show a considerable dispersion for CSPN, while the more recent results 
by  Pauldrach et al. (\citeyear{pauldrach03}; \citeyear{pauldrach04}), and Pauldrach 
(\citeyear{pauldrach05}) obtain a much lower dispersion and, consequently, a better fit to 
the line associated with  the hot, luminous stars. Our results using the N/O masses indicate 
a larger dispersion in $\log\, p_w$ for both sources of data on the stellar winds, which is 
partly a consequence of our restricted mass -- and therefore luminosity -- interval. It is 
then interesting to investigate the expected dispersion at a given luminosity, which we will 
assume to be caused by the dispersion in the stellar (and nebular) metallicity. 

The metallicity dependence of winds from red supergiants and AGB stars was recently reviewed by 
van Loon (\citeyear{vanloon}). The mass ejection process in these stars is assumed to be the 
action of the stellar radiation pressure on grains. According to van Loon (\citeyear{vanloon}), 
the dust to gas ratio $\psi$ \ is proportional to the stellar metallicity, $\psi \propto
Z$. Since the grain absorption optical depth $\tau$ \ is proportional to the dust to gas ratio, 
$\tau \propto \psi$, we have $\tau \propto Z$. It is reasonable to assume the same is true for 
CSPN, here attributing the optical depth  to the absorption in ressonance metallic lines,  so that 
we have that $p_w \propto \dot M v_\infty \propto \tau \propto Z^k$, where we may assume for 
simplicity $k \simeq 1$, combining any optical depth effects  that either $dM/dt$ \ or 
$v_\infty$ \ might have. In the case of red giants, van Loon (\citeyear{vanloon}) attributes 
the metallicity dependence on the expansion velocity, leaving the mass loss rate essentially 
independent of the metallicity. For CSPN, however, this may not apply, since the radiation 
pressure acts directly on the absorption lines, and is not a continuum-opacity process as in 
red supergiants and AGB stars.

A similar relation, namely, $p_w \propto Z^{0.9}$\ has also been suggested  for galactic hot 
stars with $\log L/L_\odot < 5.7$\ by Lamers \& Cassinelli (\citeyear{lc96}) as an upper limit 
for  the correction of the mass loss rates due to metallicity. Such a relation is similar
to the metallicity dependence of Wolf-Rayet winds, as recently studied by Vink \& de Koter 
(\citeyear{vink}) on the basis of Monte Carlo calculations. As an example, for the less evolved 
WN stars in which the wind is due to the action of the radiation pressure on many metallic lines, 
the metallicity dependence is similar as for galactic O stars, namely, $dM/dt \propto Z^m$, 
where $m \simeq 0.86$. For WC winds they find $m \simeq 0.66$.

Adopting $p_w \propto (L_\ast/L_\odot)^a$, as discussed in Section~3, where $a = 1/\alpha$, 
we can introduce a dependence on the metallicity $Z$ by writing

   \begin{equation}
     p_w \propto Z \ \left({L_\ast \over L_\odot}\right)^a \ .
     \label{pwz1}
   \end{equation}

\noindent
In the case of CSPN, the wind strength is basically related to the abundances
of highly ionized heavy elements, which produce P Cyg profiles of optical and
ultraviolet lines. Iron is the usual metallicity indicator in stars, but the 
iron lines are generally weak and the Fe abundance is depleted in photoionized
and planetary nebulae due to grain formation (cf. Perinotto et al. 
\citeyear{perinotto99}; Deetjen et al. \citeyear{deetjen}; Pottasch et al.
\citeyear{pottasch}), so that the abundances of this element measured 
in the nebulae are usually taken as lower limits. However, the metallicity 
of PN can be accurately determined on the basis of measurements of the oxygen 
abundance, as well as of other elements such as Ne, S, etc., all of which are 
strongly correlated, as shown by many recent investigations both in the Milky 
Way and in other galaxies (cf. Richer \& McCall \citeyear{richer}; 
Maciel et al. \citeyear{mci2006}).
As recently discussed by Idiart et al. (\citeyear{idiart}), average [O/Fe]
$\times$ [Fe/H] relationships can be obtained from stellar data or theoretical
models, which may provide direct O(Fe) relations. Also, recent work on the
[Fe/H] radial gradient in the galactic disk obtained from open cluster stars
and cepheid variables (Maciel et al. \citeyear{mlc2005}) shows a strong
correlation with the gradients of O, S, etc. determined from photoionized
nebulae, thus confirming these elements as representative of the nebular
metallicity. Therefore, we can safely assume that in PN and hydrogen-rich
CSPN the metal abundance $Z$ is proportional to the oxygen abundance by number O/H, 
that is, $Z = k \,$ (O/H). For example, for the solar photosphere we have 
$\epsilon({\rm O})_\odot = \log ({\rm O}/{\rm H})_\odot + 12 \simeq 8.66$ and 
$Z \simeq 0.012$ (Grevesse et al.  \citeyear{grevesse}), so that $k \simeq 26$. 
Therefore, the modified wind momentum can be written as

   \begin{equation}
    p_w \simeq \beta\ \left({L_\ast \over L_\odot}\right)^a  \ 
    \left({{\rm O} \over {\rm H}}\right)
     \label{pwz2}
   \end{equation}

\noindent
where $\beta$ is a constant to be determined. We have then

   \begin{equation}
    \log\ p_w = \log\, \beta +  a \ \log \left({L_\ast \over 
     L_\odot}\right) + \log \left({{\rm O} \over {\rm H}}\right)
     \label{pwz3}
   \end{equation}

\noindent
Using the standard definition $\epsilon({\rm O}) = \log ({\rm O}/{\rm H}) + 12$,
we have

   \begin{equation}
     \log p_w = \log\, \beta + a \ \log \left({L_\ast \over 
     L_\odot}\right) + \left[\epsilon({\rm O}) - 12\right]
     \label{pwz4}
   \end{equation}

According to M\'endez (\citeyear{mendez91}), all CSPN considered in this work are H-rich
objects, in which stellar H features are clearly distinguished in the CSPN spectrum. 
Their H and He abundances are normal, with the possible exception of NGC 2392, which shows  
some He excess. All the other objects have \lq\lq normal\rq\rq\ helium abundances
of about 10\% by number (for the actual values see M\'endez et al. \citeyear{mendez88},
\citeyear{mendez90}), again suggesting that these stars have normal Fe abundances,
and that their metal abundances are well correlated with the oxygen abundance.
Most of these stars are of spectral type Of(H), so that the He II 4686 line 
appears as a narrow emission line, and two are of type O(H), in which He II 4686 appear in 
absorption. In the sample of 115 CSPN studied by M\'endez (\citeyear{mendez91}), 62\% are H-rich
and 33\% are H-deficient. These objects have probably left the AGB with a H-rich photosphere, 
and are on their way to become DA white dwarfs. As can be seen from figure~2 of M\'endez 
(\citeyear{mendez91}), most H-rich CSPN have \lq\lq normal\rq\rq\ He abundances,
that is, He/H $\simeq 0.10$, so that their chemical composition includes about 90\% of H by 
number, as in the case of the nebulae themselves. Therefore, we can safely adopt average
oxygen abundances in these objects as in the nebulae themselves, that is

   \begin{equation}
    \epsilon({\rm O}) = \bar \epsilon({\rm PN}) = \bar \epsilon({\rm CSPN}) \simeq 8.50 \ ,
     \label{epso}
   \end{equation}

\noindent
which is essentially the same average oxygen abundances of galactic HII regions
(see for example Costa \& Maciel \citeyear{cm2006}; Henry \& Worthey \citeyear{henry}).
We can identify Eqs.~10 and 6, from which we have $a = 1.507$, and $\log \, \beta + 
[\epsilon({\rm O}) - 12] =  20.479$, so that $\log\, \beta = 23.98$.
Therefore we have

   \begin{equation}
    \log p_w = 23.98 + 1.507 \ \log \left({L_\ast \over L_\odot}\right) +
    [\epsilon({\rm O}) - 12] \ ,
    \label{pwz5}
   \end{equation}

\noindent
again with $p_w$ in cgs units (g cm s$^{-2}$). The new $p_w$ values are shown in 
the last column of Table~1. For a given luminosity, the expected 
dispersion in $\log\,p_w$ due to a difference in the metallicity is simply $\Delta \log \, 
p_w = \Delta \epsilon({\rm O})$. The two dashed lines in Fig.~3 show the results for 
$\epsilon({\rm O}) = 7.50$ and $\epsilon({\rm O}) = 9.20$, which correspond approximately to the
minimum and maximum observed values of the oxygen abundance, respectively. From the metallicity 
distribution of PN and HII regions (as well as cepheids and CSPN), we know that the large 
majority of objects have $8.00 < \epsilon({\rm O}) < 9.0$, so that the derived dispersion is 
very well defined. Table~2 shows the  dispersion $\Delta \log\,p_w$ as a function of the adopted 
value of the oxygen abundance $\epsilon({\rm O})$. It can be seen that, for the average 
$\epsilon({\rm O}) = 8.50$ and $7.5 < \epsilon({\rm O}) < 9.2$, the spread is from 
$\Delta \log\,p_w = -1.00$ to $0.70$, that is, a total spread of $1.70$ can be expected, which 
is similar to the observed scatter in Fig.~3 using the N/O masses and both the Kudritzki or 
Pauldrach databases.

\begin{table}
\caption[]{Expected dispersion of the modified wind momentum.}
\label{table2}
\begin{flushleft}
\begin{tabular}{cccc}
\noalign{\smallskip}
\hline\noalign{\smallskip}
$\epsilon_2$(O) & $\Delta \log\,p_w$ & $\epsilon_2$(O) & $\Delta \log\,p_w$\\
\noalign{\smallskip}
\hline\noalign{\smallskip}
9.20 & 0.70 & 8.30 & $-$0.20 \\
9.10 & 0.60 & 8.20 & $-$0.30 \\
9.00 & 0.50 & 8.10 & $-$0.40 \\
8.90 & 0.40 & 8.00 & $-$0.50 \\
8.80 & 0.30 & 7.90 & $-$0.60 \\
8.70 & 0.20 & 7.80 & $-$0.70 \\
8.60 & 0.10 & 7.70 & $-$0.80 \\
8.50 & 0.00 & 7.60 & $-$0.90 \\
8.40 & $-$0.10 & 7.50 & $-$1.00 \\
\noalign{\smallskip}
\hline
\end{tabular}
\end{flushleft}
\end{table}

\subsection{Empirical evidences}

Empirical evidences of metallicity effects on the modified wind momentum
of CSPN are difficult to obtain for several reasons. First, the modified
momentum probably shows a stronger dependence on the stellar luminosity,
so that any metallicity variation is expected to become apparent only for
stars having similar luminosities. Second, the actual stellar luminosity
has some intrinsic uncertainty, which basically derives from uncertainties
in the distance scale, which affects particularly planetary nebulae and
their central stars. Finally, the abundances themselves have an uncertainty
of typically 0.2 dex, depending on the element considered. Despite these
shortcomings, data on some of the objects analyzed in  this paper show
some hints in the sense that an enhanced metallicity is correlated with
a higher wind momentum. 

Taking into account the results shown in Table~1, we notice that the objects
He2-108 and He2-131 have approximately the same luminosity, $\log L/L_\odot
\simeq 3.95$. The first object is clearly more metal-poor than the second,
with a difference of about 0.5 dex, or almost a factor 3. According to
the results of Kudritzki et al. (\citeyear{kudritzki97}), which we favour here, 
He2-131 has a much higher modified wind momentum ($\log p_w \simeq 27.82$, cgs 
units) than He2-108 ($\log p_w \simeq 27.28$), which may be interpreted as a 
metallicity effect. Using our \lq\lq corrected\rq\rq\ values given in Table~1
the same pattern is observed. Also for the pair IC 4593 and Tc 1, Table~1
shows similar luminosities, namely, $\log L/L_\odot \simeq 3.77$. Again,
the first object is more metal-poor than the second by a factor of about 2.
This pattern is repeated in the determination of the modified wind momentum
by Kudritzki et al. (\citeyear{kudritzki97}) and in our corrected values given 
in Table~1, although the value by Kudritzki et al. (\citeyear{kudritzki97}) for 
Tc 1 is strictly an upper limit.

If we consider now the objects studied by Perinotto (\citeyear{perinotto}), we
notice that NGC 1535 and NGC 7009 have similar luminosities, $\log L/L_\odot 
\simeq 4.0$, but the former is more metal-poor than the latter by a factor of
1.7. Again  the same tendency is repeated in the modified wind momentum,
for which Perinotto (\citeyear{perinotto}) obtains $\log p_w \simeq 25.7$ and
$26.0$, respectively.

An indirect empirical evidence can also be obtained from the sample studied by
Malkov (\citeyear{malkov1}; \citeyear{malkov2}), in a self consistent determination
of several parameters of galactic planetary nebulae and their central stars.
Among other parameters, Malkov (\citeyear{malkov1}; \citeyear{malkov2}) obtains
central star luminosities and nebular abundances for about 130 nebulae in the
galactic disk. If we make the reasonable assumption that the modified wind
momentum $p_w$ of these objects depends on the stellar luminosity in the same
way as the small sample of CSPN with well studied winds, we could use an equation
such as eq.~(6) to estimate this quantity. We notice that 74\% of the galactic
disk nebulae in the sample have oxygen abundances higher than the average value
adopted here, $\epsilon({\rm O}) = 8.50$, while 26\% of the nebulae are metal-poor,
having $\epsilon({\rm O}) < 8.50$. Estimating the modified wind momentum from
eq.~(6), we obtain similar figures, that is, 77\% of the stars have 
$\log p_w > 25.0$ in cgs units, while 23\% have lower values of this quantity,
namely, $\log p_w < 25.0$. Of course, this is an indirect evidence that may be
affected by the assumption made and the uncertaintes involved in the determination
of these quantities, but it is reassuring that some correlation between the
modified momentum and the nebular metallicity is suggested.

\section{Summary and Conclusions}
\label{section4}

In this work we have considered a sample of well-studied CSPN for which the wind properties
as well as the basic stellar properties are known, and analyzed two different problems. 

First, we have confirmed that there is a large discrepancy between the stellar masses as derived 
from standard mass-luminosity relation for AGB stars and those derived from stellar atmosphere 
models affected by winds. We have then determined new values for the central star masses based 
on a correlation of the core stellar mass and the nebular N/O abundances, and determined the 
masses of the central stars. We showed that the masses agree very well with the independent
estimates of the masses of CSPN, and that the observed mass spread is small, entirely consistent 
with the expected mass distributions of CSPN and white dwarfs. 

Second, we have used the newly derived N/O masses and the standard mass-luminosity relation
in order to investigate the dispersion of the modified wind momentum at a given luminosity.
We have taken into account two different sets of stellar properties derived from different model
atmospheres (Kudritzki et al. \citeyear{kudritzki97}  and Pauldrach et al. \citeyear{pauldrach04}),
and showed that the dispersion on the $p_w \times L_\ast$ plane is real, and can
be well explained by the metallicity dispersion observed in the nebulae, which are a reflection
of the properties of the central stars. Similar results can also be obtained by considering the
set of objects analyzed by Perinotto (\citeyear{perinotto}), in which case the mass loss rates
have been determined with the SEI (Sobolev plus exact integration) and CFS (comoving frame)
method.

{\it Acknowledgements. This work was partly supported by FAPESP and CNPq.}

\end{document}